# Experimental Observation of Three-Component 'New Fermions' in Topological Semimetal MoP

B. Q. Lv,[1,†] Z.-L. Feng,[1,†] Q.-N. Xu,[1,†] J.-Z. Ma,[1] L.-Y. Kong,[1] P. Richard,[1,2,3] Y.-B. Huang,[4] V. N. Strocov,[5] C. Fang,[1] H.-M. Weng,[1,2] Y.-G. Shi,[1,*] T. Qian,[1,2,*] and H. Ding[1,2,3,*]

[1] *Beijing National Laboratory for Condensed Matter Physics and Institute of Physics, Chinese Academy of Sciences, Beijing 100190, China*

[2] *Collaborative Innovation Center of Quantum Matter, Beijing, China*

[3] *University of Chinese Academy of Sciences, Beijing 100190, China*

[4] *Shanghai Synchrotron Radiation Facility, Shanghai Institute of Applied Physics, Chinese Academy of Sciences, Shanghai 201204, China*

[5] *Paul Scherrer Institute, Swiss Light Source, CH-5232 Villigen PSI, Switzerland*

[†] These authors contributed to this work equally.

[*] Corresponding authors: dingh@iphy.ac.cn, tqian@iphy.ac.cn, ygshi@iphy.ac.cn

**Abstract**

Condensed matter systems can host quasiparticle excitations that are analogues to elementary particles such as Majorana, Weyl, and Dirac fermions. Recent advances in band theory have expanded the classification of fermions in crystals, and revealed crystal symmetry-protected electron excitations that have no high-energy counterparts. Here, using angle-resolved photoemission spectroscopy, we demonstrate the existence of a triply degenerate point in the electronic structure of MoP crystal, where the quasiparticle excitations are beyond the Majorana-Weyl-Dirac classification. Furthermore, we observe pairs of Weyl points in the bulk electronic structure coexisting with the 'new fermions', thus introducing a platform for studying the interplay between different types of fermions.

In quantum field theory, Lorentz invariance gives three types of fermions, namely, the Dirac, Weyl and Majorana fermions (*1,2*). While it is still under debate whether any elementary particle of Weyl or Majorana types exists, all three types of fermions have been proposed to exist as low-energy and long-wavelength quasiparticle excitations in condensed matter systems (*3-14*). The existence of Dirac and Weyl fermions has been experimentally confirmed (*15-20*) and that of Majorana fermions has been supported by various experiments (*21,22*). Recently, it has been shown theoretically that as the Poincare group (Lorentz group plus 4-translation) in the continuum space-time is reduced to the 230 space groups in lattices, more types of fermions (dubbed 'new fermions') are allowed to appear as quasiparticle excitations near certain band crossing points (*23-29*).

Specially, it is well known that fermion statistics is incompatible with three-fold degeneracy in the continuum due to the half-integer spin; yet, three-fold degeneracy (triply degenerate point (TP)) can be protected in a lattice either by rotation symmetries (*25-29*) or nonsymmorphic symmetries (*23,24*). In either case, the three-component fermions conceptually lie between Weyl fermions (two-component) and Dirac fermions (four-component) (Fig. 1A), and carry characteristic properties distinct from the other two, including unique surface states and transport features.

TPs have been recently predicted by first-principles calculations to exist in several materials with the WC-structure (*26-29*). We have performed angle-resolved photoemission spectroscopy (ARPES) measurements to investigate the electronic structure of MoP with the WC structure.

Let us first introduce the crystal and electronic structures of MoP. The space group is $P\bar{6}m2$ (No. 187). Mo and P are at the 1*d* (1/3, 2/3, 1/2) and 1*a* (0, 0, 0) Wyckoff positions, respectively. The crystal structure includes a $C_3$ rotation symmetry and a $M_z$ mirror symmetry, as illustrated in Figs. 1B and 1C. As discussed below, these space symmetries are crucial to protect the TPs.

A*b initio* calculations show a band inversion between Mo $d_{z^2}$ and $e_g$ states at the *A* point in the BZ, leading to a band crossing along Γ-*A*. Without spin-orbit coupling (SOC), the crossing bands are a non-degenerate $d_{z^2}$ orbital band and a doubly-degenerate band composed of the $e_g$ ($d_{xy}$ and $d_{x^2-y^2}$) orbitals (Fig. 1E).

The crossing point is triply degenerate and protected by the $C_3$ symmetry along $\Gamma$-$A$, which is similar to the case of the Dirac semimetals Na$_3$Bi (*7*) and Cd$_3$As$_2$ (*9*). With SOC considered, the bands along $\Gamma$-$A$ are reconstructed into two doubly-degenerate $|J_z| = 1/2$ bands and two non-degenerate $|J_z| = 3/2$ bands due to the $M_z$ mirror symmetry. The crossing points of the bands with different $|J_z|$ are protected by the $C_3$ symmetry, forming four TPs along the $\Gamma$-$A$ line (Fig. 1F).

We first perform core level photoemission measurements, which confirms the chemical composition of MoP (Fig. 2A). Figure 2 displays the overall measured electronic structures in the three-dimensional (3D) Brillouin zone (BZ). In Figs. 2C and 2D, the Fermi surfaces (FSs) and band dispersions measured in a large range of photon energy from 300 to 950 eV exhibit obvious modulations along $k_z$ with a period of $2\pi/c$. These experimental results along $k_z$ agree well with our *ab initio* calculations shown in Figs. 2B and 2D. We further measure the electronic structures in the $k_x$-$k_y$ plane with fixed photon energies. Figures 2E-2L show the measured FSs and band dispersions as well as the corresponding calculated results in the $k_z = 0$ and $\pi$ planes, respectively. We observe one hexagonal hole pocket around $\Gamma$ and one small hole pocket at $K$ at $k_z = 0$, as well as one almost circular electron pocket around $A$ at $k_z = \pi$. These experimental results in the $k_x$-$k_y$ plane are also consistent with our *ab initio* calculations.

Next we demonstrate the existence of TP in MoP by measuring the band dispersions along $\bar{\Gamma} - \bar{M}$ at different $k_z$. As indicated in Fig. 1F, four such TPs are predicted. The bands along $\Gamma$-$A$ are relatively flat around $\Gamma$ and $A$ while they disperse quickly between these two points. Due to the limitation of $k_z$ momentum resolution, the ARPES spectra of the bands with high $k_z$ dispersion are seriously broadened. It is thus hard to identify fine details in the band structure, whereas it is easy to distinguish the bands that are less dispersive along $k_z$ in ARPES experiments. Therefore, we focus on the electronic structure near the $A$ point, where it is theoretically predicted that TP #1 resides (Fig. 3L).

Figures 3A and 3B show the experimental band dispersions along $\bar{\Gamma} - \bar{M}$ at $k_z = \pi$. We observe one electron-like band and one 'W'-like band with an energy gap of 0.1 eV at $A$, which is consistent with the calculations in Fig. 3C. Note that all the bands along $\bar{\Gamma}-\bar{M}$ at $k_z = \pi$ are doubly degenerate, which is protected by the $C_{3v}$ symmetry. Upon slightly sliding the cut along $k_z$, the degeneracy is removed except for the $k_{//} = 0$ point of the electron-like band (Fig. 3F), which is protected by the $C_3$ symmetry. The splitting is observed in the 'W'-like band in Fig. 3E though they are not well separated. Upon further sliding the cut along $k_z$, the upper branch of the 'W'-like band touches the electron-like band at $k_{//} = 0$ (Fig. 3I). This touch point is triply degenerate since the electron-like band is doubly degenerate at $k_{//} = 0$ and the upper branch of the 'W'-like band is non-degenerate. In Figs. 3G and 3H, we clearly observe the bands touching at $k_{//} = 0$.

To further prove this TP, we plot the energy distribution curves (EDCs) of the ARPES intensity raw data and corresponding curvature intensity data along c3 in Figs. 3J and 3K, respectively. In the raw data (Fig. 3J), the upper branch of the 'W'-like band touches the electron-like band at $k_{//} = 0$, while the lower branch appears as a shoulder at the high binding energy side of the upper branch. We fit the EDCs of the raw data to a multiple peak function, which are exemplified in Fig. 3M. By tracking the peak positions of the fitting curves, we extract the band dispersions near the TP (Fig. 3N). In the curvature intensity data (Fig. 3K), the lower branch is clearly identified as peaks, which are well separated from the TP. These results provide direct experimental evidence of the proposed TP #1 along the $\Gamma$-$A$ line.

In addition to the TPs, we discover the existence of Weyl points (WPs) in MoP. Our calculations show a band inversion at the $K$ point. Without SOC, the band inversion forms a nodal ring centered at $K$ in the $k_z = 0$ mirror plane (Fig. 4A). With SOC included, the lack of inversion symmetry leads to Rashba-like spin splitting. The nodal rings in the mirror plane are fully gaped and pairs of WPs appear in the $\Gamma$-$K$ direction off the plane (Fig. 4A). All the WPs can be related to each other by the $D_{3h}$ crystal symmetry and thus located at the same energy. The calculated electronic structure of MoP is quite similar to that of ZrTe (*28*), except that the WPs of MoP are located at -1.1 eV, whereas those of ZrTe are at 0.05 eV above $E_F$.

In Fig. 4B, we clearly observe a ring around the $K$ point at -1.1 eV in the $k_z = 0$ plane, which is related to the calculated nodal ring without SOC. To clarify the WPs, we measure the band dispersions along $\bar{\Gamma}-\bar{K}$ at different $k_z$. The bands along

$\bar{\Gamma} - \bar{K}$ form a small gap at 0.75 $\bar{\Gamma}\bar{K}$ at $k_z = 0$ (Fig. 4H). The gap closes at $k_z = \pm 0.02$ π (Figs. 4G and 4I) and reopens at $k_z = \pm 0.05$ π (Figs. 4F and 4J). These results clearly indicate the existence of WPs along the Γ-K direction at $k_z = \pm 0.02$ π, which is consistent with the theoretical prediction (Figs. 4K-4N).

One hallmark of Weyl semimetals is Fermi arcs on the surface, which connect to the projection points of WPs with opposite chiral charges. However, we do not observe surface states related to Fermi arcs on the (001) surface, as seen from the intensity map at -1.1 eV in the $k_z = π$ plane (Fig. 4D). In the (001) surface BZ, two WPs with opposite chiral charges are projected onto the same point. It is possible that each projection point connects to itself via one arc. The arc can shrink to a point, which is equivalent to no arc at all.

Our ARPES results provide the first observation of TP in the electronic structure, demonstrating the existence of three-component 'new fermions' in MoP. As an intermediate state between Dirac and Weyl semimetals, TP topological semimetals share many similarities with them, such as a gapless Landau level under symmetry-preserving magnetic fields and topologically protected Femi arcs on certain surfaces. TP topological semimetals transit to either Weyl semimetals or nodal line semimetals under an external perturbation. One hallmark of TP topological semimetals is the FS touching near a TP, which leads to a unique spin texture and possible magnetic breakdown in quantum oscillations. All of these await experimental confirmation. The coexistence of 'new fermions' and Weyl fermions in MoP provides a platform for studying the interplay between different types of fermions, possibly offering a new route for potential applications.

**Acknowledgements:** We thank Binbin Fu, Nan Xu, and Xin Gao for the assistance in the ARPES experiments. This work was supported by the Ministry of Science and Technology of China (2016YFA0300600, 2015CB921300, 2013CB921700, 2016YFA0302400, 2016YFA0401000, and 2016YFA0300300), the National Natural Science Foundation of China (11474340, 11622435, 11234014, 11274367, 11474330,

11274362, 11674371, and 11422428), and the Chinese Academy of Sciences (XDB07000000). Y.B.H. acknowledges support by the CAS Pioneer Hundred Talents Program. The ARPES experiments were performed at the Dreamline beam line of the Shanghai Synchrotron Radiation Facility and at the ADRESS beam line of the Swiss Light Source.

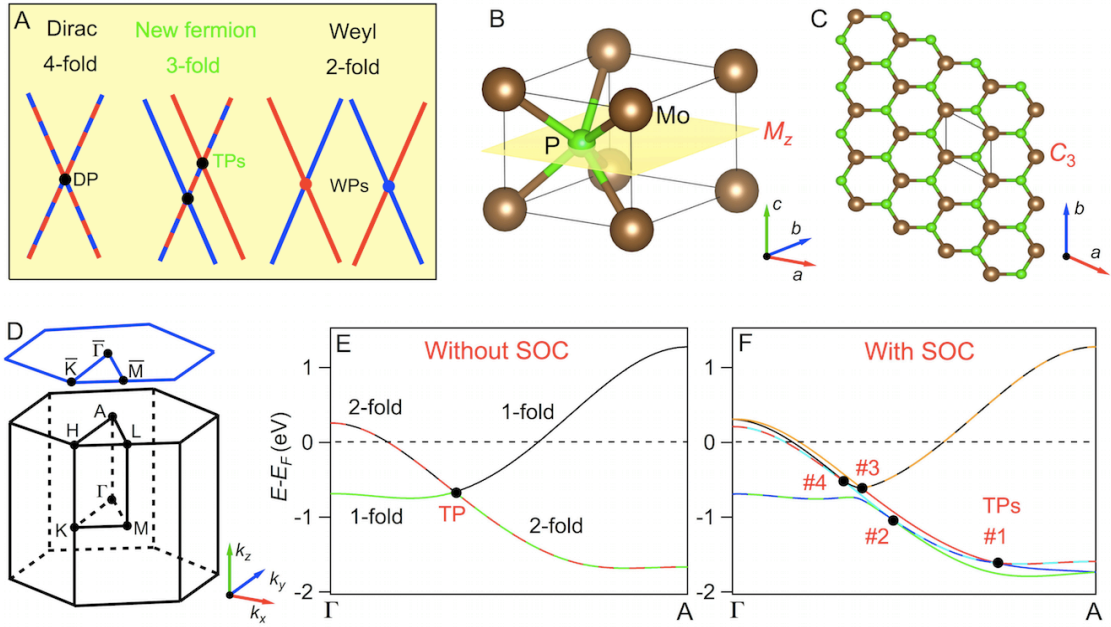

Fig. 1. Crystal structure and band structure along Γ-A of MoP. (A) Schematic of band crossing points with 4-fold, 3-fold, and 2-fold degeneracy, which are denoted as DP, TP, and WP, respectively. (B) 3D crystal structure of MoP in one unit cell. The yellow plane indicates the mirror plane $M_z$. (C) Top view of the lattice showing the $C_3$ rotation symmetry. (D) 3D bulk BZ as well as projected (001) surface BZ with high-symmetry points indicated. (E) and (F) Calculated band structures along Γ-A without and with SOC, respectively. The solid circles at the crossing points indicate the TPs. The lines with mixed colors represent doubly degenerate bands. The lines with pure color represent non-degenerate bands.

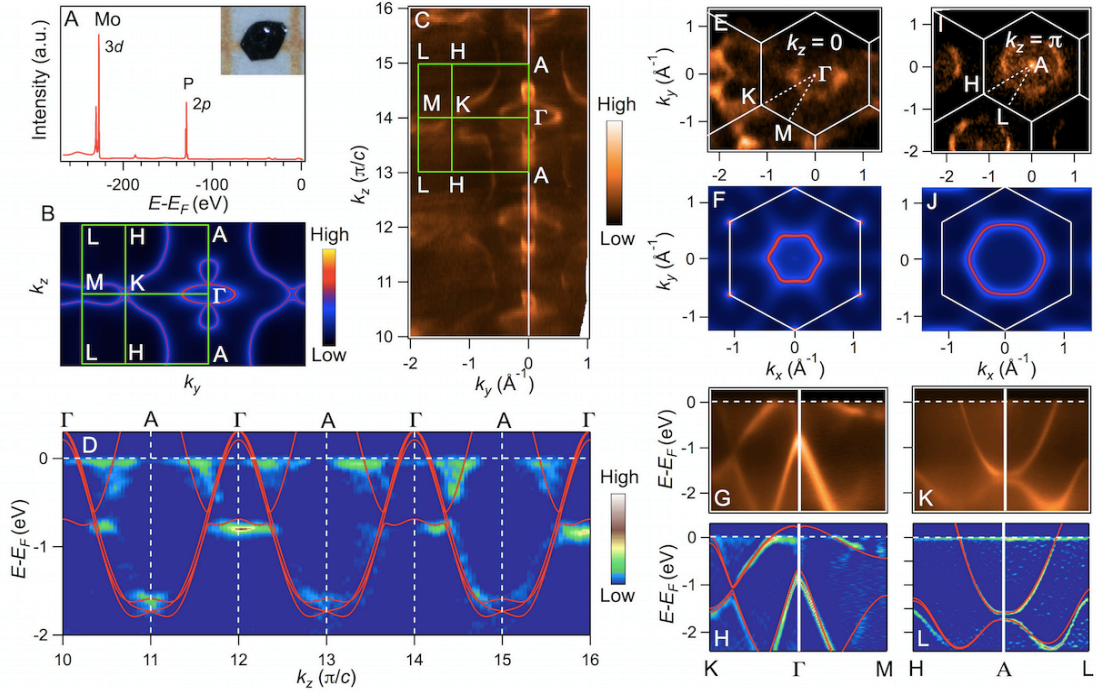

Fig. 2. Overall electronic structure of MoP in the 3D bulk BZ. (A) Core level photoemission spectrum showing characteristic Mo 3*d* and P 2*p* peaks. The inset shows a typical single crystal of MoP. (B) and (C) Calculated and experimental FSs in the Γ-*K*-*M*-*L*-*H*-*A* plane, respectively. (D) Curvature intensity map showing band dispersions along Γ-*A*. For comparison, the calculated band structure along Γ-*A* is overlapped on top of the experimental data. (E) and (F) Experimental and calculated FSs in the $k_z = 0$ plane, respectively. (G) and (H) ARPES intensity plot and curvature intensity plot, respectively, showing band dispersions along *K*-Γ-*M*. For comparison, the calculated band structure along *K*-Γ-*M* is overlapped on top of the experimental data in (H). (I) to (L) Same as (E) to (H) but in the $k_z = \pi$ plane. The ARPES results at $k_z = 0$ and $\pi$ are taken at photon energy $h\nu = 435$ and $510$ eV, respectively.

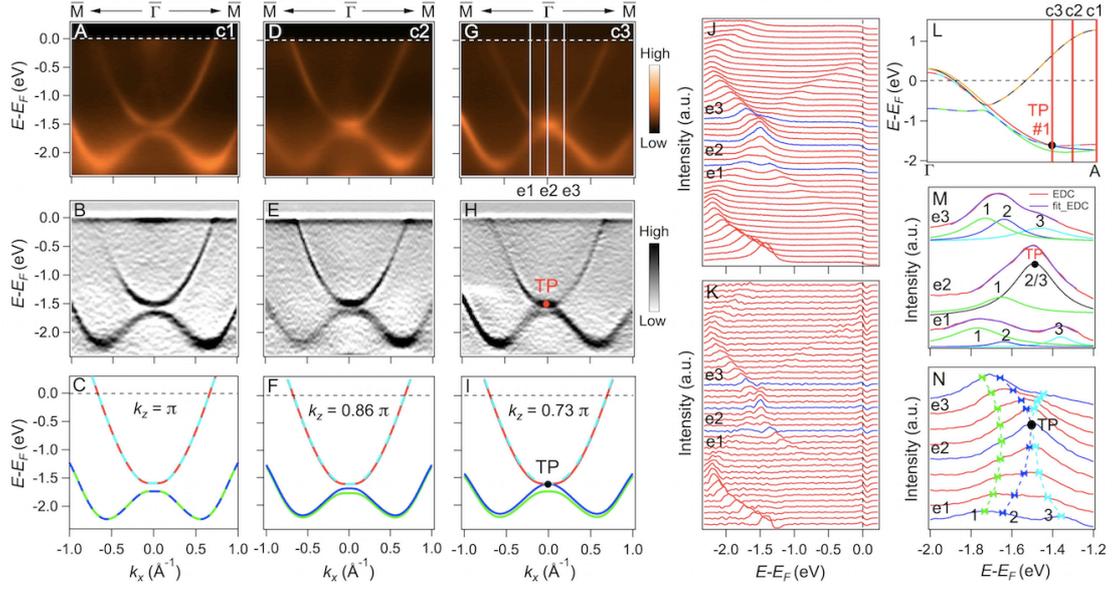

Fig. 3. Electronic structure near the TP #1. (A) to (C) ARPES intensity plot, curvature intensity plot, and calculated band structure along $\bar{\Gamma} - \bar{M}$ at $k_z = \pi$, respectively. (D) to (F) Same as (A) to (C) but at $k_z = 0.86\,\pi$. (G) to (I) Same as (A) to (C) but at $k_z = 0.73\,\pi$. (J) and (K) EDC plots of the ARPES intensity in (G) and the curvature intensity in (H), respectively. The EDCs with blue color in (J) and (K) correspond to the vertical lines e1, e2, and e3 in (G). (L) Calculated band structure along $\Gamma$-$A$. The vertical lines c1, c2, and c3 indicate the $k_z$ locations of the cuts in (A to C), (D to F), and (G to I), respectively. (M) Three representative EDCs e1, e2, and e3 fitted to a multiple Lorenz peak function. (N) EDCs of the ARPES intensity near the TP #1 in (G). The symbols indicate the extracted band dispersions by tracking the peak positions of the fitting curves. Peaks #1 and #2 represent the lower and upper branches of the 'W'-like band, respectively. Peak #3 represents the electron-like band, where the splitting is not distinguished in the experimental data since the energy scale is very small, as seen from the calculated band structure in (I).

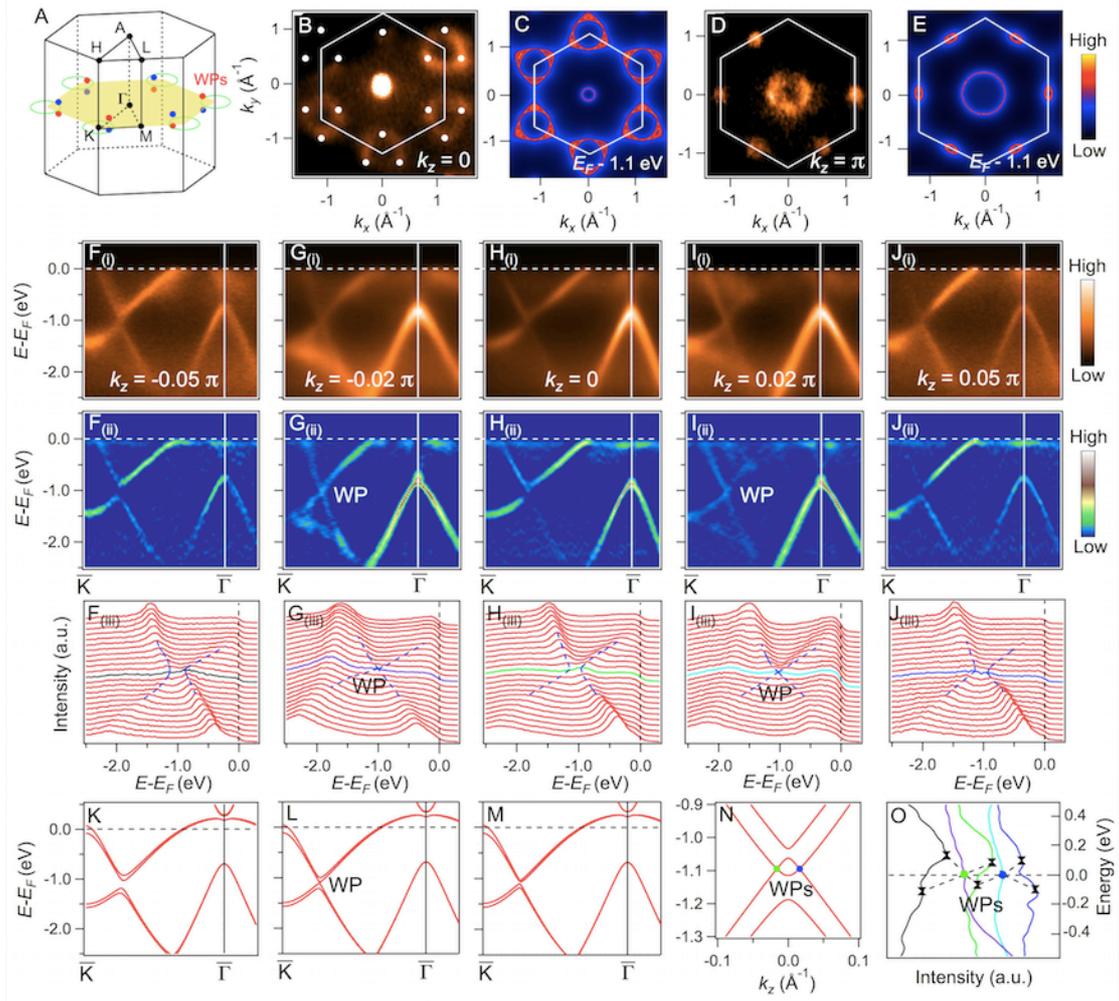

Fig. 4. Electronic structure near the WPs. (A) 3D bulk BZ with nodal lines without SOC in the $k_z = 0$ mirror plane and WPs with SOC off the plane. (B) and (C) Experimental and calculated intensity plots at -1.1 eV in the $k_z = 0$ plane, respectively. White dots in (B) indicate the projected locations of the WPs in the plane. (D) and (E) Same as (B) and (C) but in the $k_z = \pi$ plane. (F) (i to iii) ARPES intensity plot, curvature intensity plot, and EDCs near the WP along $\bar{\Gamma}-\bar{K}$ at $k_z = -0.05\,\pi$, respectively. (G to J) (i to iii) Same as (F) but at $k_z = -0.02\,\pi$, 0, 0.02 $\pi$, and 0.05 $\pi$, respectively. (K) to (M) Calculated band structures along $\bar{\Gamma}-\bar{K}$ at $k_z = \pm 0.05\,\pi$, $\pm 0.02\,\pi$, and 0, respectively. (N) Calculated band structure along $k_z$ through one pair of WPs. (O) EDCs along $k_z$ through one pair of WPs, which correspond to the EDCs with black, purple, green, light blue, and blue colors in (F to J) (iii), respectively. Zero energy in (O) is set to the energy of the WPs. Green and blue solid circles indicate the locations of the bulk Weyl nodes with opposite chiral charges.